\documentclass[sn-mathphys-num]{sn-jnl}% Math and Physical Sciences Numbered Reference Style 
\usepackage{graphicx}%
\usepackage{multirow}%
\usepackage{amsmath,amssymb,amsfonts}%
\usepackage{amsthm}%
\usepackage{mathrsfs}%
\usepackage[title]{appendix}%
\usepackage{xcolor}%
\usepackage{textcomp}%
\usepackage{manyfoot}%
\usepackage{booktabs}%
\usepackage{algorithm}%
\usepackage{algorithmicx}%
\usepackage{algpseudocode}%
\usepackage{listings}%

\usepackage{lscape}
\theoremstyle{thmstyleone}%
\theoremstyle{thmstyletwo}%

\theoremstyle{thmstylethree}%

\begin{document}

\title[GRAMEP, an alignment-free method based on the Principle of Maximum Entropy]{GRAMEP: an alignment-free method based on the Maximum Entropy Principle for identifying SNPs}


\author[1]{\fnm{Matheus Henrique} \sur{Pimenta-Zanon}}\email{omatheuspimenta@outlook.com}

\author[1]{\fnm{André Yoshiaki} \sur{Kashiwabara}}\email{kashiwabara@utfpr.edu.br}

\author[2]{\fnm{André Luís Laforga} \sur{Vanzela}}\email{andrevanzela@uel.br}

\author*[1]{\fnm{Fabricio Martins} \sur{Lopes}}\email{fabricio@utfpr.edu.br}

\affil[1]{\orgdiv{\orgdiv{Computer Science Department}, \orgname{Universidade Tecnol\'ogica Federal do Paran\'a (UTFPR)}, \orgaddress{\street{Alberto Carazzai, 1640}, \postcode{86300-000}, \state{Corn\'elio Proc\'opio - Paraná}, \country{Brazil}}}}

\affil[2]{\orgdiv{Laboratory of Cytogenetics and Plant Diversity, Department of General Biology}, \orgname{Universidade Estadual de Londrina}, \orgaddress{\street{Rodovia Celso Garcia Cid, PR-445, Km 380}, \postcode{86057-970}, \state{Londrina - Paraná}, \country{Brazil}}}


\abstract{
%The Abstract should not exceed 350 words. Please minimize the use of abbreviations and do not cite references in the abstract. The abstract must include the following separate sections:

\textbf{Background:} Advances in high throughput sequencing technologies provide a huge number of genomes to be analyzed. Thus, computational methods play a crucial role in analyzing and extracting knowledge from the data generated. Investigating genomic mutations is critical because of their impact on chromosomal evolution, genetic disorders, and diseases. It is common to adopt aligning sequences for analyzing genomic variations. However, this approach can be computationally expensive and restrictive in scenarios with large datasets.\\
\textbf{Results:} We present a novel method for identifying single nucleotide polymorphisms (SNPs) in DNA sequences from assembled genomes. This study proposes GRAMEP, an alignment-free approach that adopts the principle of maximum entropy to discover the most informative k-mers specific to a genome or set of sequences under investigation. The informative k-mers enable the detection of variant-specific mutations in comparison to a reference genome or other set of sequences. In addition, our method offers the possibility of classifying novel sequences with no need for organism-specific information. GRAMEP demonstrated high accuracy in both in silico simulations and analyses of viral genomes, including Dengue, HIV, and SARS-CoV-2. Our approach maintained accurate SARS-CoV-2 variant identification while demonstrating a lower computational cost compared to methods with the same purpose. \\
\textbf{Conclusions:} GRAMEP is an open and user-friendly software based on maximum entropy that provides an efficient alignment-free approach to identifying and classifying unique genomic subsequences and SNPs with high accuracy, offering advantages over comparative methods. The instructions for use, applicability, and usability of GRAMEP are open access at \url{https://github.com/omatheuspimenta/GRAMEP.} }




\maketitle

\section{Background}\label{sec_background}

% The Background section should explain the relevant context and the specific issue that the software described is intended to address.

The analysis of genomic sequences has been extensively studied to understand species' diversity and evolution~\cite{zielezinski_alignment-free_2017}. With the advancement of sequencing technology, an ever-increasing amount of genomics data is being generated, opening up new perspectives for research on genetic variants in various organisms \cite{de_pierri_sweep_2020}.

The genome of a species encodes the instructions required for the production of thousands of proteins and RNA molecules \cite{crick_central_1970}. This information is embedded within the DNA sequence, acting as a complex code. Mutations in nucleotides, the DNA building blocks, can be linked to typos in this code, altering DNA sequences and, consequently, the sequence of RNA and synthesized proteins. These variations can affect the organism’s phenotype or its observable characteristics. Mutations that affect only a single base pair are known as single nucleotide variants (SNVs) or single nucleotide polymorphisms (SNPs). SNPs are akin to minute variations in the genetic code and can have varying effects on the organism. Some SNPs may be silent, causing no significant changes. Others may have a mild impact, while some can lead to drastic alterations in the phenotype, such as genetic diseases \cite{snustad_principles_2015}.

SNVs can lead to the simultaneous emergence of different phenotypes and result in intraspecific variations. In viruses, such as SARS-CoV-2, several mutations occurred during the 2019 pandemic. Depending on the specific mutations, these variations directly affected the transmission rate, mortality, and infectivity of the virus \cite{tian_emergence_2022,worobey_dissecting_2021,perico_genomic_2022,franceschi_mutation_2021,harvey_sars-cov-2_2021}. The study of genomic variation is critical for diagnosing, preventing, and treating diseases. When applied to the study of viral diseases, genomic variation analysis is relevant for epidemiological purposes and for controlling viral spread. Additionally, a systematic understanding of the evolution and taxonomy of various species, including viruses, relies on genomic variation research \cite{andersen_proximal_2020}. Furthermore, organisms like the Dengue virus (DENV) exhibit weak error correction mechanisms, leading to a high mutation rate and significant diversity within their variant populations \cite{bock_evolutionary_2006}. The presence of SNPs in viral genomes can have various consequences, such as changes in resistance to antiviral drugs (e.g., Influenza virus and HCV) \cite{bock_evolutionary_2006} or enabling immune system evasion (e.g., human immunodeficiency virus type 1 (HIV-1), however, a high mutation rate can also lead to the extinction of certain variants \cite{yeo_determination_2020,cuevas_extremely_2015}.

The precise identification of SNVs in a large number of sequences can be computationally challenging, considering classical alignment methods. The complexity can be polynomial regarding the number and size of the analyzed sequences \cite{lange_mathematical_2002}. Alignment-based approaches are widely used in biological sequence analysis like BLAST \cite{altschul_basic_1990} to identify regions of similarity between sequences. However, they face significant challenges when dealing with multi-genomic scale data because of computational resources. For instance, considering the existence of gaps, the number of possible alignments for two sequences with $100$ base pairs can reach an order of approximately $10^60$. Finding the best alignment based on a scoring system that provides a value for match, mismatch, and gap penalties is a well-known problem that dynamic programming algorithms, such as Smith-Waterman \cite{smith1981identification} and Needleman-Wunsch \cite{needleman_general_1970}, can solve in quadratic time complexity. However, the unfeasibility of these alignment-based approaches for large-scale analysis due to the time complexity \cite{zielezinski_alignment-free_2017,zielezinski_benchmarking_2019} underscores further research for alignment-free methods.

It is well known that the high mutation rates are characteristic of viruses and represent a challenge for traditional sequence analysis methods, which rely on aligning sequences to identify similarities \cite{zielezinski_alignment-free_2017}. Alignment-free approaches offer an essential alternative for analyzing these highly mutable genomes, mainly because they do not rely on finding regions with high identity \cite{zielezinski_benchmarking_2019}. Researchers have adapted alignment-free methods for comparative analyses \cite{zielezinski_alignment-free_2017,zielezinski_benchmarking_2019} with many distinct datasets. Furthermore, alignment-free methods are often scalable and computationally efficient, making them a suitable alternative for handling large datasets \cite{forsdyke2019success}. Unlike alignment-based approaches, they do not assume collinearity between sequences—meaning they do not require a one-to-one correspondence of residues along the sequence—making them ideal for capturing complex genetic patterns in viruses that undergo frequent recombination \cite{solis2018open}, horizontal gene transfer \cite{cong2016novel},  duplications \cite{song2008sequence}, and gene losses \cite{duffy_rates_2008}.

In this context, heuristics and alignment-free methods emerge as a viable alternative, providing near-optimal solutions in feasible time and allowing scalability in the analysis of large volumes of data \cite{zielezinski_alignment-free_2017,zielezinski_benchmarking_2019}. They offer a more efficient approach for the analysis of complete genomes with lower computational complexity, which is achieved by considering mathematical and computational concepts, such as calculus, information theory, statistics, physics, and linear algebra being more adaptable to different types of sequences, especially those with high mutation rates, such as those found in viruses \cite{zhang_review_2017,murugaiah_novel_2021,zielezinski_benchmarking_2019}. Alignment-free methods have applications in broad areas within computational biology, such as a study of viral diversity, identification of genes and regulatory regions, analysis of phylogeny and molecular evolution, and detection of genetic variants associated with diseases \cite{marchet_blight_2021,ito_basinetbiological_2018,de_pierri_sweep_2020,amin_evaluation_2019}.

Genomic sequence analysis can be conducted using two primary alignment-free approaches: word-based and information theory-based methodologies \cite{zielezinski_alignment-free_2017}. Word-based methods focus on analyzing k-mers, short subsequences of fixed-length \emph{k} extracted from the sequences under study. These k-mers are treated as distinct units, and their frequency and distribution within genomic sequences are used to generate profiles and sequence-specific patterns. Conversely, information theory-based methods employ mathematical and statistical tools from information theory to quantify the information content across genomic sequences. This approach enables the identification of complex, contextual patterns within sequences, aiding in the detection of functional and evolutionary relationships among different genomic regions \cite{barros-carvalho_efficient_2017, vopson_dynamics_2022, vopson_new_2021}.

Several methods adopt the spectrum generated by k-mer occurrence frequency \cite{kuksa_efficient_2009}. This technique is based on the premise that different species exhibit unique k-mer patterns, facilitating the identification and classification of sequences according to their k-mer frequency \cite{chor_genomic_2009}. Alternatively, recent approaches have explored the use of genetic algorithms to identify deterministic subsequences of interest \cite{fiscon_missel_2016}.

Incorporating deterministic subsequences as features offers the potential to reduce dimensionality when training machine learning models \cite{lebatteux_machine_2022, lebatteux_machine_2024, lebatteux_combining_2021, lebatteux_toward_2019, fiscon_missel_2016}. This approach leverages these subsequences as primary features to characterize input sequences. Additionally, machine learning models trained on such features can develop the ability to recognize complex and subtle patterns, ultimately enabling the classification of novel sequences into distinct species.

Thanos et al. \cite{thanos_entropic_2018} introduces a genomic analysis approach based on information theory, specifically utilizing Shannon entropy applied to non-overlapping blocks of subsequences. This framework adopts Shannon entropy to detect repetitive regions, which often show significant entropy fluctuations compared to other genomic segments. This feature enables the identification of repetitive elements, such as transposons and duplicated sequences, which play critical roles in evolution and genetic regulation. 

The GENIES method \cite{vopson_new_2021, vopson_dynamics_2022} offers a distinct approach to mutation detection in viral genomes, with a particular emphasis on SARS-CoV-2. It leverages the entropy spectrum, a graphical representation of the relationship between block index and k-mer entropy. Unlike the method proposed by Thanos, GENIES utilizes overlapping subsequences with a calculated step size to evaluate entropy for each k-mer within the genome. Mutations are identified by comparing the entropy ratios of corresponding k-mer blocks between the reference and variant sequences; deviations from a ratio of 1 indicate the presence of mutations within specific k-mers. While this technique is effective for pinpointing specific mutations, a key limitation is its requirement for equal-length reference and variant sequences, rendering GENIES inapplicable to studies involving viral genomes of varying lengths.

MEME (discriminative mode) \cite{bailey_value_2010,bailey_meme_2015} employs a statistical sequence model based on user-defined parameters, if available, for expected site count and width. It allows the incorporation of prior sequence information and offers two search options: zero or one occurrence per sequence (ZOOPS) and one occurrence per sequence (OOPS). STREME \cite{bailey_streme_2021} utilizes a generalized suffix tree and evaluates motifs using a unilateral statistical test of enrichment within a specified sequence set compared to a control set.

Lebatteux et al. \cite{lebatteux_toward_2019} introduced the CASTOR-KRFE method to identify discriminative subsequences within viral genomes for classification. This method employs a feature selection to identify the most informative k-mers, subsequently using them as features for the classification of clades, species, or sub-variants. An evolution of CASTOR-KRFE, the KEVOLVE method \cite{lebatteux_combining_2021,lebatteux_machine_2024}, retains this structure but replaces the feature selector with a genetic algorithm for k-mer selection. The latest refinement, KANALYZER \cite{lebatteux_kanalyzer_2022}, leverages these discriminative k-mers to pinpoint specific gene regions containing mutations within the analyzed genomes. It accomplishes this by aligning only in regions where discrepancies occur between the matches of discriminative k-mers and those of the reference sequence. This approach enables the identification of single nucleotide polymorphisms (SNPs) and insertions/deletions (indels), which may impact amino acid changes.

In light of this scenario, we propose a new method leveraging the principle of maximum entropy to pinpoint the most informative deterministic regions unique to each species, clade, or sub-variant within an organism using the k-mers approach. These regions subsequently serve as features for training a robust biological sequence classification model. The proposed approach enables the identification of SNPs within the analyzed organism’s genome, providing scalability, allowing vertical expansion through increased hardware capacity or horizontal distribution across multiple processing nodes, ensuring its feasibility and effectiveness for extensive genomic analysis. We show that our approach can analyze the genomes of RNA viruses like SARS-COV, DENV, and HIV \cite{olson_introducing_2023}, which are known for their high recombination rates and reassignment rates. GRAMEP (Genome Variation Analysis from the Maximum Entropy) is an open software that encompasses its key functionalities in genome analysis.

\section{Implementation}\label{sec_implementation}

% This should include a description of the overall architecture of the software implementation, along with details of any critical issues and how they were addressed.

\begin{figure}[h]
\centering%
\includegraphics[width=1\linewidth]{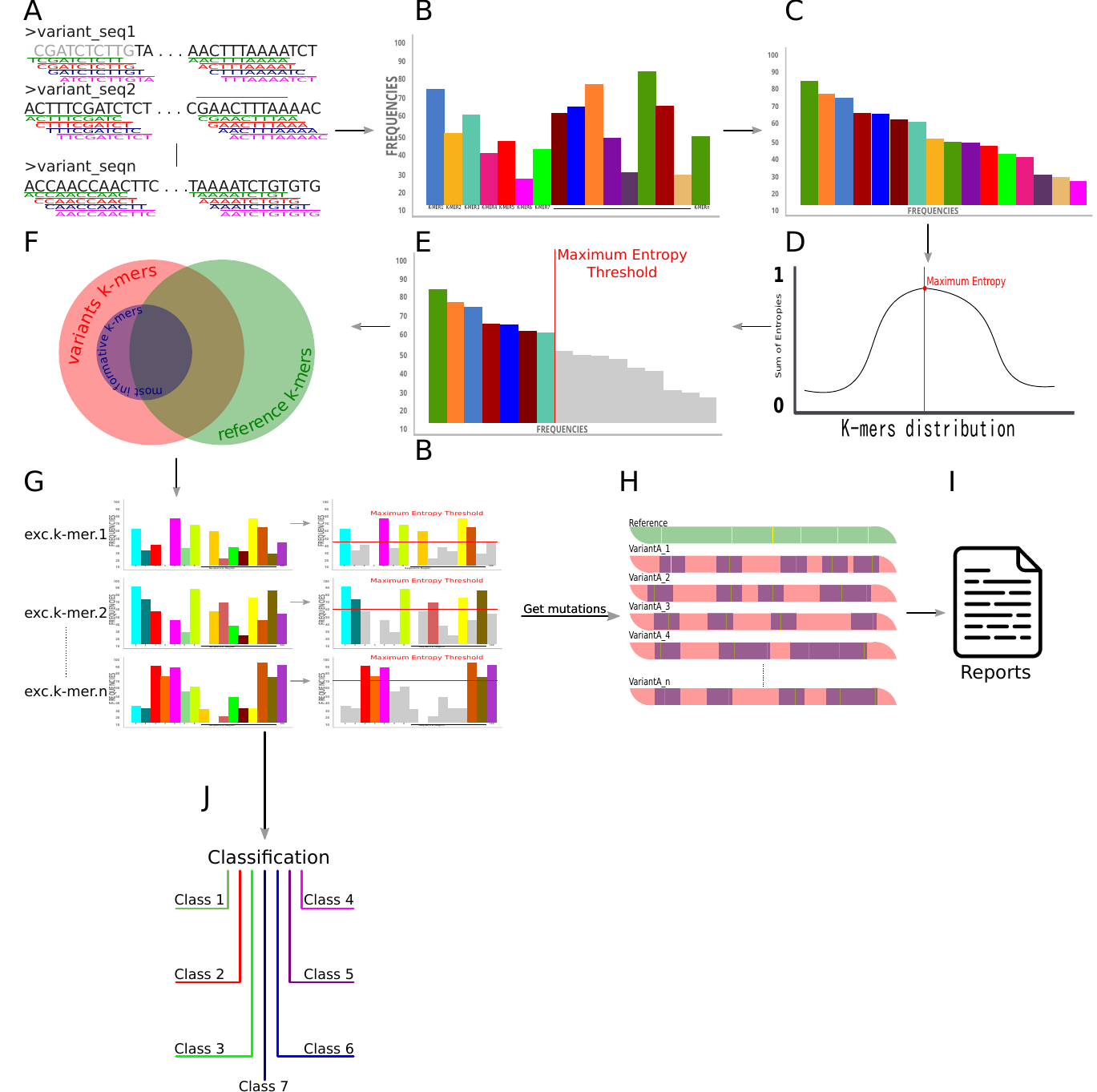}
\caption{Overview of the GRAMEP method.}%
\label{fig:overview}
\end{figure}

GRAMEP - Genome vaRiation Analysis from the Maximum Entropy, is written mostly in Python with some functions written in Rust. Figure \ref{fig:overview} gives an overview of the method.

\subsection{Data input and initial processing}

The \texttt{get-mutations} function requires two fasta files: one containing a single reference sequence and another containing the sequences of the variant under analysis. To identify exclusive subsequences, the function uses two primary parameters: `word' and `step'.

The `word' parameter determines the k-mer size, defining the length of the subsequences to be considered during the analysis. The `step' parameter, on the other hand, sets the offset for the sliding window, specifying how far the window moves as it traverses the sequences. These parameters directly influence the size and overlap of the extracted subsequences, allowing flexibility in the selection process according to the analysis needs.

For instance, Figure \ref{fig:overview} - Panel A illustrates the extraction process of $10$-length k-mers with a step size of $1$, applied to the variant sequences. This configuration generates overlapping subsequences of length $10$ across the sequence, providing a detailed view of the variation.

In addition to the variant and reference sequences, the k-mer size (`word'), and the step (`step'), it is necessary to specify the location where the analysis results will be saved by using the `save-path' parameter. Other optional parameters can be set to adjust the analysis as needed. These optional parameters include the annotation file (.gff), allowing mutations to be associated with specific genomic regions such as genes or exons, thereby facilitating the biological interpretation of the results displayed in reports at the end of the execution; the maximum number of allowed SNP mutations per k-mer, which sets the limit of single nucleotide polymorphisms (SNPs) permitted for each k-mer; the type of sequence being analyzed, enabling specification of whether the sequences are DNA, RNA, or other types, which influences the preprocessing of sequences during loading. The loading process removes sequences containing characters incompatible with the specified sequence type. 

Furthermore, there is the option to generate a complete mutation report, and finally, the chunk size, a parameter that defines the number of sequences to be processed simultaneously in each block (chunk), optimizing parallel analysis. This value depends on the available memory, balancing the efficient use of computational resources with the system's capacity.
These additional configurations increase the flexibility and precision of the analysis, allowing customization of parameters according to the objectives and technical limitations of the computational environment.

While loading sequences, each sequence is divided into $100$ non-overlapping regions to obtain the occurrence frequency of k-mers in each of these regions. In addition to recording the occurrence frequency of each k-mer, this information is stored in a hashmap, which also includes the occurrence frequency of the k-mer in each of the $100$ regions. In this way, the hashmap is structured so that the keys correspond to the k-mers, while the values represent both the total occurrence frequency of the k-mer and the occurrence frequency in each specific region.

After loading the occurrence frequencies of the k-mers and the regions, the next step involves calculating the maximum entropy. This calculation is essential for identifying which k-mers and regions are most informative, thus enabling a data-driven selection of the k-mers and regions that hold greater relevance for the analysis at hand.

\subsubsection{Entropy}
The entropy permeates diverse scientific domains, encompassing both microscopic and macroscopic scales. In statistical mechanics \cite{tsallis_nonadditive_2009}, it reflects the disorder within a system. Similarly, thermodynamics leverages entropy to describe energy exchange and equilibrium states \cite{clausius_mechanical_2019}. While Clausius and Kelvin established the second law of thermodynamics, the underlying formula for entropy remained elusive until Boltzmann and Gibbs' groundbreaking work \cite{boltzmann_theoretical_2012}. Their microscopic approach defined the now-renowned Boltzmann-Gibbs entropy:

\begin{equation}
\label{eq:gibbs_entropy}
S = -k \sum_{i=1}^{W} p_i \log p_i
\end{equation}
\noindent where $k$ is the Boltzmann constant and $p_i$ represents the probabilities of the $W$ possible microscopic states $\displaystyle \sum_{i=1}^{W} p_i = 1$.

Meanwhile, information theory employs a distinct concept of entropy introduced by Shannon \cite{shannon_mathematical_1948}. Here, entropy $H(X)$ quantifies the uncertainty associated with a random variable $X$, formally defined as:

\begin{equation}
\label{eq:shannon_entropy}
H(X) = - \sum_{x \in \mathcal{X}} p(x) \log p(x)
\end{equation}
\noindent where $x$ represents each element in the alphabet of $\mathcal{X}$ and $p(x)$ denotes its probability. Notably, entropy solely depends on the probabilities, not the specific values, of $X$.

Jaynes \cite{jaynes_information_1957-1} bridged these realms by connecting the thermodynamic entropy of Boltzmann and Gibbs with Shannon's information-theoretic entropy. The principle of maximum entropy posits that, given available data, one should choose the probability distribution with the highest possible entropy. This essentially implies utilizing only the observed data without incorporating any prior assumptions.

Consider a discrete distribution with $n$ events and observed frequencies $h_1, h_2, \ldots, h_n$. Let $p_i = \displaystyle\frac{h_i}{N}$ (where $N$ is the total number of samples) represent the probability of the i-th outcome. For a distribution with two classes, $A$ and $B$, their respective entropies are:

\begin{equation}
H(A) = - \sum_{i=1}^{s} \frac{p_i}{P_A} \log \left( \frac{p_i}{P_A} \right),
\label{eq:MEA}
\end{equation}

\begin{equation}
H(B) = - \sum_{i=s+1}^{n} \frac{p_i}{P_B} \log \left( \frac{p_i}{P_B} \right),
\label{eq:MEB}
\end{equation}
where $P_A$ and $P_B$ are the respective probabilities of belonging to class $A$ or $B$ $(P_A + P_B = 1)$. By maximizing the sum of these class entropies $H(A)+H(B)$, we achieve the maximum entropy $(ME)$:
\begin{equation}
\label{eq:ME}
ME = \mathop{\arg \max}\limits_{s = 1,2,\ldots,n} \{H(A)+H(B)\}.
\end{equation}
This maximization essentially identifies the point in the distribution where the classes are most separable, signifying maximum uncertainty between them \cite{guiasu_principle_1985}. Leveraging maximum entropy to distinguish class distributions \cite{kapur_new_1985} allows us to focus solely on data with high information content, effectively filtering out noise and bias. This approach proves particularly valuable in high-dimensional problems, where it can significantly reduce dimensionality \cite{barros-carvalho_efficient_2017}.

\subsection{Selection of the most informative k-mers and regions}

This step of the GRAMEP begins by sorting the k-mer frequencies in descending order, generating a frequency histogram, Panel C of Figure \ref{fig:overview} illustrates this step. Subsequently, occurrence probabilities are calculated using the maximum entropy principle. Following this, an automatic threshold for ``informative'' subsequences is established. Panels D and E of Figure \ref{fig:overview} graphically illustrate the obtaining of the maximum entropy value and also the selection of the most informative k-mers from the histogram distribution. K-mers surpassing this threshold are then considered informative and retained to constitute the variant-specific set of informative subsequences. The entropy calculation is dependent on the data used as input, which can influence the choice of the most informative k-mers in each class, as the maximum entropy principle solely considers the probability of event occurrences. This reliance on probability precludes intermediate values from influencing the maximum entropy calculation. Therefore, utilizing multiple sequences per variant is recommended to determine the optimal cutoff point between these classes.

Following the identification of informative k-mers specific to the variant, this step accomplishes this by subtracting the set of informative k-mers belonging to the variant from the comprehensive set of k-mers present in the organism's reference sequence, Panel F of Figure \ref{fig:overview} illustrates through Venn diagrams representing the set of k-mers from the reference sequence in green, the set of k-mers from the variant in red and the subset of the most informative k-mers for the variant in blue. This subtraction process effectively filters out k-mers shared between the variant and the reference, ensuring the resulting set solely comprises unique and informative subsequences characteristic of the analyzed variant.

Algorithm \ref{alg:firststep} provides a summary of how the most informative exclusive k-mers are obtained for each variant under analysis.
\begin{algorithm}
\caption{Obtaining the most informative exclusive k-mers.}\label{alg:firststep}
\begin{algorithmic}[1]
\Require{Reference.FASTA, Variant.FASTA, word, step}
\Ensure{Most informative exclusive k-mers}
\For{sequence in Variant.FASTA sequences}
        \State Get the frequencies of the k-mers and regions;
      \EndFor
\State Sort the frequencies of occurrence of k-mers in descending order\;
\State Get the maximum entropy from the frequencies of occurrence\;
\State Apply the automatic cutoff (threshold) to the k-mers\;
\State Extract the k-mers from the reference\;
\State Get the most informative exclusive k-mers for the variant\;
\end{algorithmic}
\end{algorithm}

The next step involves identifying the most informative regions where each previously selected k-mer appears within the sequences. To accomplish this, each sequence is divided into $100$ regions, and for each most informative exclusive k-mer of the variant under analysis, a threshold is determined based on the k-mer's frequency in these regions. This procedure is performed similarly to the k-mer selection step, employing the principle of maximum entropy, as illustrated in panel G of Figure \ref{fig:overview}.

\subsection{Obtaining mutations and reports}

The next step focuses on detecting mutations and their corresponding positions within the reference genome. This task leverages the established set of exclusive and informative subsequences associated with each variant sequence and their positions. Only subsequences with potential mutation sites within the variant sequence are selected for further analysis. This selection is achieved by considering the intersection of the variant's exclusive subsequences, effectively filtering out non-mutated regions. 

The selection of regions relative to the reference sequence where the mutation search will take place is performed by considering the possible variations in size and positions between the reference sequence and the variant sequence. First, the variation between the length of the reference sequence and the variant is calculated. If the variation is significant, we calculate an extra adjustment. Then, we use this adjustment to expand the start and end points of each region, creating a wider range to account for the variation. %If this variation exceeds $1\%$, then the additional position is defined as $\text{additional\_position} = \left\lfloor \frac{\text{sequence\_variation}}{0.01} \right\rfloor$, otherwise, the additional position is set to $1$. The range for each region is then determined by \([ \text{start\_value}, \text{end\_value} ]\), where $\text{start\_value} = (\text{selected\_index} \times 0.01) \times \text{reference\_length} - \text{additional\_position}$ and $\text{end\_value} = ((\text{selected\_index} + 1) \times 0.01) \times \text{reference\_length} + \text{additional\_position}$.

These bounds are always adjusted to ensure they do not exceed the length of the reference sequence. This approach allows only the most informative regions to be analyzed, avoiding the processing of the entire sequence. As a result, the strategy reduces the occurrence of false positives by focusing the analysis on areas most relevant (discriminant) to the comparison between the reference and the variant.

Subsequently, for each of these mutation-prone subsequences, the Levenshtein distance is adopted \cite{levenshtein_binary_1966}. Panel H of Figure \ref{fig:overview} illustrates this step. Each most informative unique k-mer and their position is represented in purple in the variant sequences, and SNPs are identified in yellow.

The following step culminates with a comprehensive analysis of mutation abundance and distribution within each variant sequence (Panels H and I in Figure \ref{fig:overview}). This step starts with the computation of a frequency table, systematically tallying the occurrence and position of every identified mutation. Furthermore, if feasible, a graphical representation is generated, visually depicting the entirety of mutation locations relative to the reference genome.
In the presence of an annotation file in .gff3 format, detailed mutation information can be generated for each sequence. This enriched data encompasses sequence identification, functional annotation, start and end positions of the mutation, type of mutated region, both variant and reference subsequences harboring the mutation, and specific nucleotide changes. Analyzing multiple variants of the same organism enables the identification of shared mutations, offering valuable insights into population-level trends and evolutionary trajectories.

\subsection{Classification and prediction}

The exclusive and informative k-mers sets for each variant, obtained in the previous steps, are combined to form a single set containing all unique k-mers for each variant on the reference genome, using the \texttt{classify} function. If the most informative exclusive k-mers are not available, they can be obtained via the `get-kmers' parameter. Subsequently, these k-mers are adopted as features to train a classification and prediction model capable of effectively classifying novel sequences.

The next step involves training the classification and prediction model. This step utilizes each k-mer in the consolidated set as a feature, resulting in a feature matrix, as shown in Panel J in Figure \ref{fig:overview}. Each row in this matrix represents a sequence, while each column corresponds to the occurrence frequency of a specific k-mer within that sequence. Following the acquisition of raw data (i.e., occurrence frequencies). The Min-Max rescaling is employed to improve usability in the model. Following data processing, a Random Forest algorithm with default parameters is implemented for training and classification tasks. A ``10-fold'' cross-validation approach is utilized to validate the generated model, demonstrate its robustness, and mitigate overfitting during the training step.

\section{Results and discussion}\label{sec_results}

% This should include the findings of the study, including, if appropriate, results of statistical analysis, which must be included either in the text or as tables and figures. This section may be combined with the Discussion section for Software articles.
\subsection{GRAMEP - identification of SNPs mutations}

To assess the GRAMEP for accurate SNPs identification, simulations were performed by considering datasets based on HIV and DENV genomes, motivated by other studies \cite{struck_comet_2014,lebatteux_kanalyzer_2022}. Simulated mutations incorporated parameters reflecting real-world variation, including sequence length (based on average lengths in available datasets), virus-specific mutation rates, sequencing error rates, and genome size variation (based on standard deviation sequence length in available datasets). For each scenario, a fixed-size reference sequence was generated, followed by the creation of $1,000$ mutated sequences incorporating size variations, sequencing errors, and true mutations. 

The adopted parameters for HIV-1 (S-HIV) and DENV (S-DENV) simulations were obtained from literature \cite{plummer_dengue_2015,bock_evolutionary_2006,cuevas_extremely_2015,yeo_determination_2020}. The S-HIV dataset consisted of strings with a length of $8,981$, a mutation rate of $3 \times 10^{-3}$, and a variation rate of $0.0222$. These parameters were based on available sequences in the Los Alamos Sequence Database (\url{https://www.hiv.lanl.gov/}). The S-DENV dataset, on the other hand, comprised strings with a length of $10,553$, a mutation rate of $1 \times 10^{-3}$, and a variation rate of $0.0205$. Both simulations employed a uniform error rate of $5 \times 10^{-4}$, and the GRAMEP parameters used for simulations were obtained from empirical experiments presented in Supplementary Material - Appendix \ref{ap:parameters}. In this case, we used the ``word'' and ``step'' values of $15$ and $1$, respectively. 

The simulations were performed considering the parameter configuration specified above. Notably, the method achieved a false positive rate (FPR) of zero in both simulations, demonstrating that all identified mutations were confirmed as true positives. Additionally, by providing the frequency of mutation occurrences, GRAMEP allows users to easily check if any mutations have a high occurrence frequency, thereby facilitating detailed analysis of relevant mutations, excluding point mutations arising from sequencing errors or individual sample peculiarities.

Regarding the true positive rate, the GRAMEP achieves performance exceeding $93\%$ in both simulated scenarios. This result indicates a high capacity of the method to identify most of the present mutations, which is further supported by the low rate of false negatives. Furthermore, when evaluating classification metrics such as accuracy, Matthews correlation coefficient (MCC), and F1-score, the effectiveness of GRAMEP in accurately classifying mutations is evident, reinforcing its applicability for genomic variant analysis. Further evaluation metrics are provided in Table \ref{tab:results_simulation}.

\begin{table}[!h]
\centering
\caption{Average and (standard deviation) obtained from performing 1,000 simulations on {\it in silico} data. Adopted metrics: (TPR) True Positive Rate, (FPR) False Positive Rate, (TNR) True Negative Rate, (FNR) False Negative Rate, (ACC) accuracy, (MCC) Matthews Correlation Coefficient and F1-score.}
\label{tab:results_simulation}
\begin{tabular}{|l|l|l|l|l|l|l|l|}
\hline
       & TPR                                                    & FPR                                                 & TNR                                                   & FNR                                                   & ACC                                                    & MCC                                                    & F1                                                     \\ \hline
S-HIV  & \begin{tabular}[c]{@{}l@{}}93.85\\ (7.53)\end{tabular} & \begin{tabular}[c]{@{}l@{}}0.0\\ (0.0)\end{tabular} & \begin{tabular}[c]{@{}l@{}}99.99\\ (0.0)\end{tabular} & \begin{tabular}[c]{@{}l@{}}6.14\\ (7.53)\end{tabular} & \begin{tabular}[c]{@{}l@{}}96.73\\ (2.43)\end{tabular} & \begin{tabular}[c]{@{}l@{}}93.75\\ (4.53)\end{tabular} & \begin{tabular}[c]{@{}l@{}}96.66\\ (4.22)\end{tabular} \\ \hline
S-DENV & \begin{tabular}[c]{@{}l@{}}93.70\\ (7.55)\end{tabular} & \begin{tabular}[c]{@{}l@{}}0.0\\ (0.0)\end{tabular} & \begin{tabular}[c]{@{}l@{}}99.99\\ (0.0)\end{tabular} & \begin{tabular}[c]{@{}l@{}}6.30\\ (7.55)\end{tabular} & \begin{tabular}[c]{@{}l@{}}96.92\\ (3.76)\end{tabular} & \begin{tabular}[c]{@{}l@{}}94.24\\ (6.91)\end{tabular} & \begin{tabular}[c]{@{}l@{}}96.58\\ (4.21)\end{tabular} \\ \hline
\end{tabular}
\end{table}

In order to assess GRAMEP in the identification of SNPs in real scenarios, it was adopted a dataset of $20$ SARS-CoV-2 strains encompassing $579,053$ genomic sequences. The dataset of the SARS-CoV-2 virus was extracted from NCBI. The $20$ lineages with the highest number of available sequences until January 2024 were selected. The reference genome used was also obtained from NCBI, corresponding to the Wuhan (identification NC\_045512.2). Details were presented in Table \ref{tab:sars-cov-2}.

\begin{table}[!ht]
\centering
\caption{SARS-CoV-2 adopted dataset to assess the \texttt{get-mutations} function.\label{tab:sars-cov-2}}
\label{tab:sars-cov-2}
\begin{tabular}{lc}
\multicolumn{2}{c}{\textbf{SARS-CoV-2}}                                                     \\ \hline
\multicolumn{1}{|l|}{\textbf{Lineages}} & \multicolumn{1}{c|}{\textbf{Number of sequences}} \\ \hline
\multicolumn{1}{|l|}{AY.103}            & \multicolumn{1}{c|}{33,450}                       \\ \hline
\multicolumn{1}{|l|}{AY.25}             & \multicolumn{1}{c|}{12,961}                       \\ \hline
\multicolumn{1}{|l|}{AY.3}              & \multicolumn{1}{c|}{14,237}                       \\ \hline
\multicolumn{1}{|l|}{AY.44}             & \multicolumn{1}{c|}{19,658}                       \\ \hline
\multicolumn{1}{|l|}{B.1.526}           & \multicolumn{1}{c|}{12,725}                       \\ \hline
\multicolumn{1}{|l|}{B.1}               & \multicolumn{1}{c|}{12,478}                       \\ \hline
\multicolumn{1}{|l|}{B.1.1.7}           & \multicolumn{1}{c|}{65,875}                       \\ \hline
\multicolumn{1}{|l|}{BA.1.15}           & \multicolumn{1}{c|}{20,153}                       \\ \hline
\multicolumn{1}{|l|}{BA.1.18}           & \multicolumn{1}{c|}{13,342}                       \\ \hline
\multicolumn{1}{|l|}{BA.1.1}            & \multicolumn{1}{c|}{89,471}                       \\ \hline
\multicolumn{1}{|l|}{B.1.2}             & \multicolumn{1}{c|}{29,619}                       \\ \hline
\multicolumn{1}{|l|}{BA.1}              & \multicolumn{1}{c|}{15,182}                       \\ \hline
\multicolumn{1}{|l|}{BA.2.12.1}         & \multicolumn{1}{c|}{76,893}                       \\ \hline
\multicolumn{1}{|l|}{BA.2}              & \multicolumn{1}{c|}{49,668}                       \\ \hline
\multicolumn{1}{|l|}{BA.4.6}            & \multicolumn{1}{c|}{12,751}                       \\ \hline
\multicolumn{1}{|l|}{BA.5.1}            & \multicolumn{1}{c|}{14,064}                       \\ \hline
\multicolumn{1}{|l|}{BA.5.2.1}          & \multicolumn{1}{c|}{34,914}                       \\ \hline
\multicolumn{1}{|l|}{BA.5.2}            & \multicolumn{1}{c|}{17,994}                       \\ \hline
\multicolumn{1}{|l|}{BA.5.5}            & \multicolumn{1}{c|}{16,868}                       \\ \hline
\multicolumn{1}{|l|}{BQ.1.1}            & \multicolumn{1}{c|}{16,750}                       \\ \hline
\multicolumn{1}{|l|}{\textit{Total}}    & \multicolumn{1}{c|}{579,053}                      \\ \hline
\end{tabular}
\end{table}

For validation, GRAMEP was compared against data from COV2Var \cite{feng_cov2var_2024}, a resource analyzing and annotating mutations in over 13 billion SARS-CoV-2 sequences sourced from GISAID. The ``word'' and ``step'' parameters were set to $15$ and $1$. The large number of sequences employed served to demonstrate the scalability of the proposed methodology. In addition, it demonstrates the methodology's scalability, allowing thousands of sequences to be analyzed simultaneously.

To assess the effectiveness of GRAMEP in identifying structural mutations in each variant, results were analyzed based on three different sequence cut-offs (thresholds): $90\%$, $95\%$, and $99\%$ of sequences processed by both COV2Var and GRAMEP. Table \ref{tab:results_cov2var} presents the obtained results, indicating the percentage of mutations identified by GRAMEP that are also present in COV2Var. It can be observed that as the cut-off increases, the concordance between mutations detected by GRAMEP and those cataloged by COV2Var for SARS-CoV-2 variants also increases.

When considering only mutations present in $99\%$ of the analyzed sequences in each variant, the overlap between mutations recorded by COV2Var and those detected by GRAMEP approaches $100\%$ across the evaluated variants, except for two variants. These results indicate that GRAMEP is effective in identifying recurrent structural mutations at high-frequency levels.

The mutations identified by GRAMEP exhibited high concordance with those present in $99\%$ of the sequences belonging to the COV2Var-analyzed variants (based on GISAID \cite{khare_gisaids_2021} data). The identified mutations represented, on average, more than $98\%$ of all variants found in $99\%$ of the sequences. Moreover, none of the identified SNP-type mutations were false positives, further reinforcing the ability to pinpoint mutations truly present in the majority of analyzed sequences. All outputs of the get-mutations function are detailed in the supplementary materials.

An additional potential application of GRAMEP involves identifying mutations common to specific organism variants. This capability stems from the fact that all identified mutations inherently belong to the examined variant. In this context, we explored shared mutations among the SARS-CoV-2 variants, using \texttt{get-intersections} function. For example, mutation C14408T was present in all analyzed variants, leading to a proline-to-leucine substitution at amino acid position 4715 \cite{pachetti_emerging_2020}. A mutation at position A23403G within the Spike (S) protein region has been observed, leading to the substitution of aspartic acid (D) with glycine (G) at this residue. The remaining identified mutations were predominantly present in most analyzed sequences, suggesting distinctive variations between the variants. Additionally, GRAMEP offers the capability to generate detailed reports for each analyzed sequence, outlining the identified mutations, which facilitates interpretation and comparison of results.

\begin{table}[!h]
\centering
\caption{Mutations found for each variant after running GRAMEP on the SARS-CoV-2 dataset compared to the mutations found in 90\%, 95\%, and 99\% of the sequences of the COV2Var database and GRAMEP results.}
\label{tab:results_cov2var}
\begin{tabular}{|l|c|c|c|}
\hline
Variant                                             & GRAMEP (90\%)                                          & GRAMEP (95\%)                                          & GRAMEP (99\%)                                          \\ \hline
AY.3                                                & 94.28                                                  & 93.10                                                  & 100                                                    \\ \hline
AY.25                                               & 94.11                                                  & 100                                                    & 100                                                    \\ \hline
AY.44                                               & 93.75                                                  & 95.65                                                  & 100                                                    \\ \hline
AY.103                                              & 94.11                                                  & 96.00                                                  & 100                                                    \\ \hline
B.1                                                 & 100                                                    & 100                                                    & 100                                                    \\ \hline
B.1.1.7                                             & 78.12                                                  & 88.46                                                  & 100                                                    \\ \hline
B.1.526                                             & 86.66                                                  & 85.71                                                  & 100                                                    \\ \hline
B.1.2                                               & 100                                                    & 100                                                    & 100                                                    \\ \hline
BA.1                                                & 89.74                                                  & 96.55                                                  & 100                                                    \\ \hline
BA.1.1                                              & 88.09                                                  & 96.77                                                  & 100                                                    \\ \hline
BA.1.15                                             & 88.88                                                  & 97.14                                                  & 100                                                    \\ \hline
BA.1.18                                             & 90.00                                                  & 92.59                                                  & 90.90                                                  \\ \hline
BA.2                                                & 87.93                                                  & 91.11                                                  & 100                                                    \\ \hline
BA.2.12.2                                           & 88.73                                                  & 87.71                                                  & 100                                                    \\ \hline
BA.4.6                                              & 85.52                                                  & 88.40                                                  & 84.21                                                  \\ \hline
BA.5.1                                              & 90.62                                                  & 91.30                                                  & 100                                                    \\ \hline
BA.5.2                                              & 91.30                                                  & 92.72                                                  & 100                                                    \\ \hline
BA.5.2.1                                            & 92.64                                                  & 92.72                                                  & 100                                                    \\ \hline
BA.5.5                                              & 89.55                                                  & 89.47                                                  & 100                                                    \\ \hline
BQ.1.1                                              & 90.66                                                  & 96.15                                                  & 100                                                    \\ \hline
\begin{tabular}[c]{@{}l@{}}Mean\\ (SD)\end{tabular} & \begin{tabular}[c]{@{}l@{}}90.73\\ (4.73)\end{tabular} & \begin{tabular}[c]{@{}l@{}}93.58\\ (4.19)\end{tabular} & \begin{tabular}[c]{@{}l@{}}98.75\\ (3.98)\end{tabular} \\ \hline
\end{tabular}
\end{table}

Given its data-driven nature, GRAMEP is well-suited for application across diverse scenarios and organisms. This is because it solely relies on information extracted from the analyzed sequences, eliminating the need for prior knowledge about the organisms under investigation.

\subsection{GRAMEP - classification and prediction of sequences}
The GRAMEP extends beyond mutation identification and offers the potential for biological sequence classification and prediction. This can be achieved by leveraging the most informative exclusive k-mers associated with each variant within an organism.

To demonstrate this application, a dataset of DENV genomes containing four dengue serotypes was obtained through BV-BRC \cite{olson_introducing_2023}. The reference genome adopted for the Dengue virus was extracted from NCBI (identification NC\_001477.1). 

\begin{table}[!ht]
\centering
\caption{Dataset used to execute experiment to asses the GRAMEP \texttt{classify} function.\label{tab:denv}}
\label{tab:denv}
\begin{tabular}{lc}
\multicolumn{2}{c}{\textbf{Dengue Virus (DENV)}}                                            \\ \hline
\multicolumn{1}{|l|}{\textbf{Serotype}} & \multicolumn{1}{c|}{\textbf{Number of sequences}} \\ \hline
\multicolumn{1}{|l|}{Type 1}            & \multicolumn{1}{c|}{2,571}                        \\ \hline
\multicolumn{1}{|l|}{Type 2}            & \multicolumn{1}{c|}{1,756}                        \\ \hline
\multicolumn{1}{|l|}{Type 3}            & \multicolumn{1}{c|}{1,272}                        \\ \hline
\multicolumn{1}{|l|}{Type 4}            & \multicolumn{1}{c|}{462}                          \\ \hline
\multicolumn{1}{|l|}{\textit{Total}}    & \multicolumn{1}{c|}{6,061}                        \\ \hline
\end{tabular}
\end{table}

To evaluate the performance of the GRAMEP in terms of classification, the same methodology of repeated K-fold cross-validation described in \cite{lebatteux_machine_2024} was applied. In this process, 100 training datasets were created, each consisting of 250 sequences randomly selected from each serotype, totaling 1000 sequences in the training set. The sequences not selected for the training set comprise the test set, totaling 5051 sequences, ensuring that the model is evaluated on previously unused data. This approach allows for a robust analysis of the method's ability to accurately classify the sequences, ensuring variability in the samples and enhancing the generalization of the results.

The ``word'' parameter was set to $9$, and the ``step'' parameter was assigned a value of $1$, aligning with the values used in other methods analyzed to ensure consistency across the analysis. Following training and validation on dedicated datasets, we employed the trained models for prediction on independent test data.

\begin{figure}[h]
\centering%
\includegraphics[width=1\linewidth]{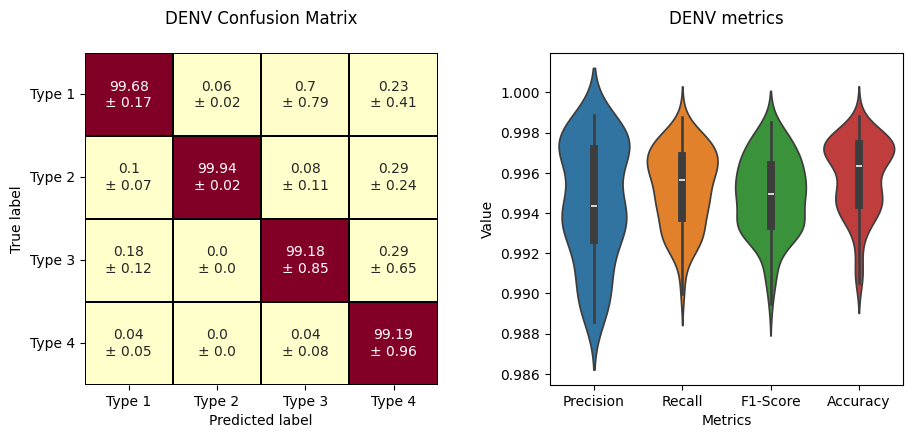}
\caption{Confusion matrix and metrics from DENV lineages prediction from GRAMEP.}%
\label{fig:conf_mtx_denv}
\end{figure}

To assess the performance of the classification models, the standard metrics, including precision, recall, F1-score, Matthews correlation coefficient (MCC), and accuracy, were adopted. For clarity and comprehensiveness, the confusion matrix and metrics for DENV are shown in Figure \ref{fig:conf_mtx_denv}.

GRAMEP offers a reduction in computational complexity compared to alignment methods. This advantage stems from the requirement of solely analyzing k-mer occurrence frequencies within the sequences. Consequently, our method facilitates the simultaneous analysis of large datasets, even on personal computers.

\subsection{Comparing GRAMEP to existing state-of-the-art tools}

To a broader comparison of the results of the proposed approach with similar methods of classification and identification of mutations from deterministic regions in the genome in the literature, it was carried out a review and four methods commonly used and studied in problems similar to those presented previously.

The performance of several motif discovery tools in identifying discriminatory sequence regions within SARS-CoV-2 genomes representing different variants was considered. Important methods such as MEME suite (MEME) \cite{bailey_value_2010,bailey_meme_2015}, STREME \cite{bailey_streme_2021}, CASTOR-KRFE \cite{lebatteux_toward_2019}, and KEVOLVE \cite{lebatteux_machine_2024}, were considered to compare their results from the previous work by \cite{lebatteux_machine_2024}.

\begin{figure}[!htb]
\centering%
\includegraphics[width=1\linewidth]{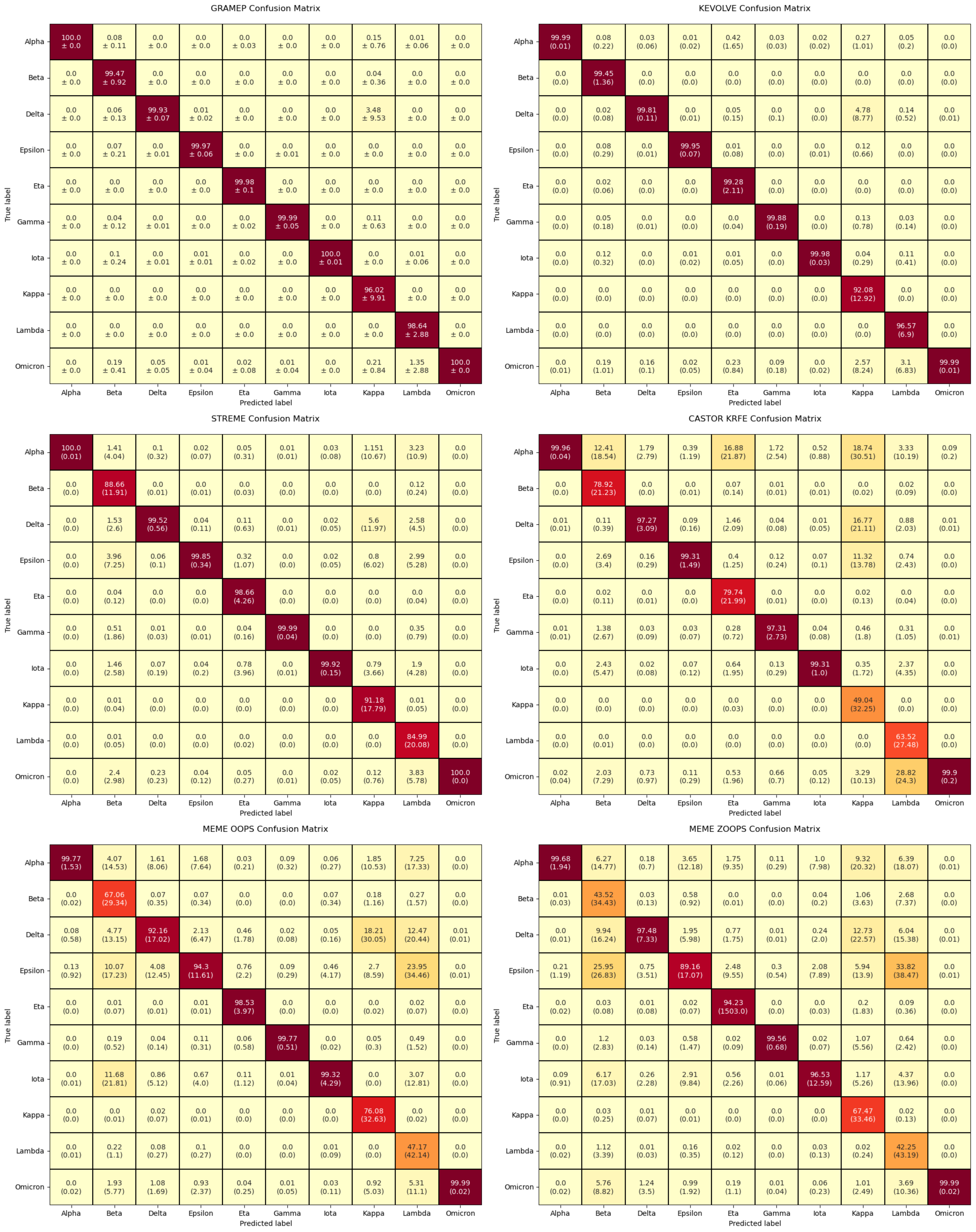}
\caption{Results of the comparative study.}%
\label{fig:conf_mtx_compare}
\end{figure}

A dataset of $334,956$ SARS-CoV-2 genomes representing ten World Health Organization (WHO) cataloged variants was analyzed, as shown in Table \ref{tab:kevolve_dataset}. K-fold cross-validation was performed with a random selection of sequences in $100$ iterations. Each fold comprised $2,500$ sequences for training and the remainder for testing. To ensure balanced representation within the training sets despite varying numbers of available sequences per variant, $250$ sequences were allocated to each variant except Kappa $(100)$, Alpha $(350)$, and Omicron $(300)$. Each tool was used to identify discriminatory motifs within the training sets, which were subsequently utilized to train a machine-learning algorithm using the k-mer and step size equal $9$ and $1$, respectively. We compare the performance of GRAMEP to that achieved by \cite{lebatteux_machine_2024} using Figure \ref{fig:conf_mtx_compare}, which presents the confusion matrices considering the prediction accuracy for each variant class.

\begin{table}[!ht]
\caption{KEVOLVE dataset~\cite{lebatteux_machine_2024}. \label{tab:kevolve_dataset}}
\centering
\begin{tabular}{|ll|l|}
\hline
\multicolumn{1}{|l|}{WHO Labe} & Pango Lineage   & Number of sequences \\ \hline
\multicolumn{1}{|l|}{Alpha}    & B.1.1.7         & 175,212             \\ \hline
\multicolumn{1}{|l|}{Beta}     & B.1.351         & 695                 \\ \hline
\multicolumn{1}{|l|}{Gamma}    & P.1             & 8,129               \\ \hline
\multicolumn{1}{|l|}{Delta}    & B.1.617.2       & 9,408               \\ \hline
\multicolumn{1}{|l|}{Kappa}    & B.1.617.1       & 127                 \\ \hline
\multicolumn{1}{|l|}{Epsilon}  & B.1.427/B.1.429 & 14,674              \\ \hline
\multicolumn{1}{|l|}{Iota}     & B.1.526         & 19,274              \\ \hline
\multicolumn{1}{|l|}{Eta}      & B.1.525         & 716                 \\ \hline
\multicolumn{1}{|l|}{Lambda}   & C.37            & 428                 \\ \hline
\multicolumn{1}{|l|}{Omicron}  & B.1.1.529/BA.x  & 106,293             \\ \hline
\multicolumn{2}{|l|}{Total number of sequences}  & 334,956             \\ \hline
\end{tabular}
\end{table}

%We compare the performance of GRAMEP to that achieved by \cite{lebatteux_machine_2024} using Figure \ref{fig:parallel_compare}, which presents the metrics achieved by each method.

\begin{figure}[!ht]
\centering%
\includegraphics[width=1\linewidth]{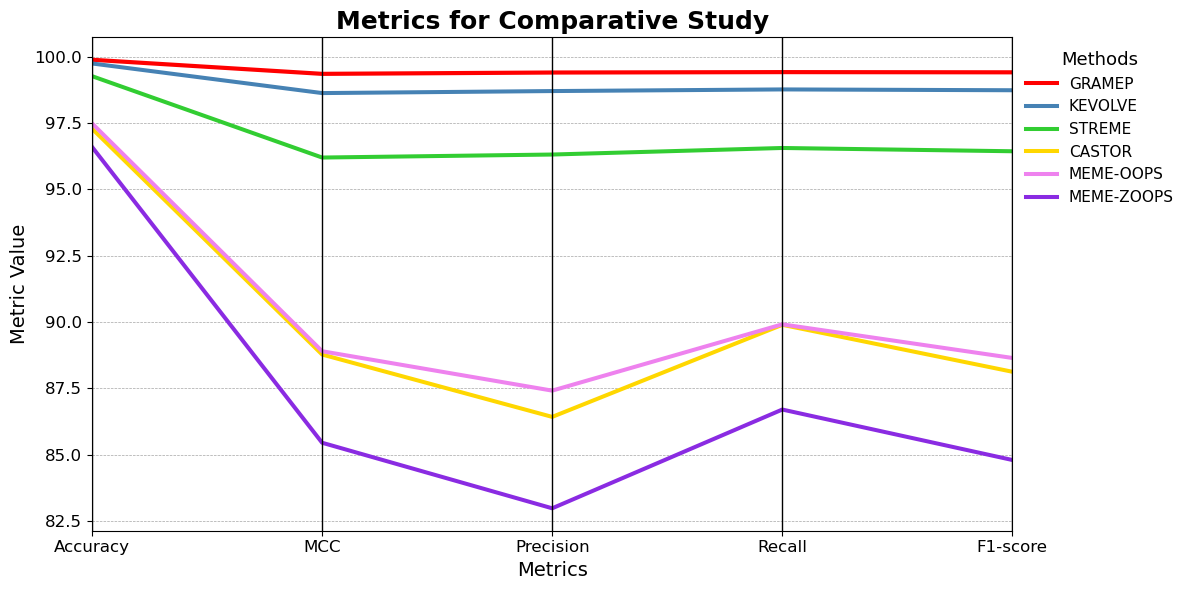}
\caption{Confusion matrix from DENV lineages prediction.}%
\label{fig:parallel_compare}
\end{figure}

The analysis of the obtained metrics, alongside the confusion matrix, highlights the GRAMEP's ability to accurately classify new sequences based on a relatively small training dataset. This outcome suggests the method’s generalization and indicates that the sub-regions extracted for each variant are indeed discriminative for their respective classes.

The metrics achieved values close to $100\%$, outperforming the competitor methods, as confirmed by the confusion matrix analysis. Specifically examining the confusion matrix generated by GRAMEP for ten SARS-CoV-2 variant classes, only two classes, Kappa and Lambda, showed values slightly below $99\%$, with averages of $96\%$ and $98\%$, respectively. This trend of lower accuracy in these variant classes was also observed with the other methods, though with a higher rate of misclassification. In particular, the Kappa variant showed some incorrect predictions relative to the Delta variant, which is reasonable given the high degree of genetic similarity between these variants.

As a comparative result, GRAMEP outperformed the other methods, indicating its assertiveness and suitability, offering a contribution as a method and open software for the use and replication of the present study.

\section{Conclusion}\label{sec_conclusion}

% This should state clearly the main conclusions and provide an explanation of the importance and relevance of the case, data, opinion, database or software reported.

Identifying and classifying mutations within genomes are crucial tasks underpinning advancements in public health research, including drug and vaccine development, disease control strategies, and various other areas. However, these tasks present significant challenges. Traditional methods often rely on sequence alignment, which can be computationally expensive, require specific reference information, and potentially generate inaccurate results because of genomic sequences' inherent complexity and variability. Consequently, alignment-free approaches have emerged as a promising alternative for mutation identification. While various methods have been proposed, each employs its unique approach.

This study presents a novel method, called GRAMEP, for selecting the most informative subsequences from genomic data, grounded in the principle of maximum entropy. This approach leverages information theory, particularly Shannon entropy, to identify k-mers that are most discriminative for characterizing different variants of an organism. By prioritizing these informative subsequences, we propose to create unique k-mer signatures for each variant.

Beyond simply selecting the most informative k-mers, the GRAMEP application demonstrates the methodology's potential in genome analysis. The four scenarios analyzed concern obtaining SNPs in silico and viral organisms, identifying mutations in common between variants of the same organism, and classifying sequences from different organisms.

The proposed method is able to discover SNPs with a high-reliability rate and also obtain exclusive regions for each variant of the reference, which can be applied as ``barcodes'' for classifying these organisms. Unlike other methods designed with similar objectives and functionalities, such as KEVOLVE and CASTOR-KRFE, GRAMEP offers a key advantage in requiring only the sequences of the variant under analysis to identify and extract the most discriminative sub-regions. In contrast, tools like KEVOLVE and CASTOR-KRFE generally require data from multiple variants to perform a comparative analysis and identify distinctive regions. This feature enables GRAMEP to be applied more efficiently and with less reliance on external data, making it particularly useful in contexts where access to multiple variants may be limited.

In terms of classification, using an automatic threshold based on maximum entropy reduces the dimensionality of the feature space, maintaining satisfactory accuracy. In addition, our methodology does not depend on prior information; only the input sequences are considered to extract the discrimination features, making it a useful tool in a variety of different scenarios and organisms.

Currently, the GRAMEP primarily focuses on identifying single nucleotide polymorphisms (SNPs) and excludes mutations like insertions or deletions. As a result, real-world applications, such as the SARS-CoV-2 analysis, solely report identified SNP mutations. Future updates could explore this approach further to optimize parametrization for different scenarios. Beyond basic k-mer frequencies, the threshold derived from the maximum entropy principle could be applied to other features extracted from the sequences. These features could encompass physical-chemical and/or biological properties relevant to the specific organism under analysis.

GRAMEP method was implemented in open source and is available at GitHub: \url{https://github.com/omatheuspimenta/GRAMEP} under the open-source MIT license. GRAMEP documentation is available from Read the Docs: \url{https://gramep.readthedocs.io/en/latest/}. The random sequence generator used during in silico simulations is available at Github: \url{https://github.com/omatheuspimenta/seq_generatoRS}.

\bmhead{Acknowledgements}
This study was financed by the Coordenação de Aperfeiçoamento de Pessoal de Nível Superior (CAPES) - Finance Code 001, the Fundação Araucária (Grant number 035/2019, 138/2021 and NAPI - Bioinformática) and CNPq (Grant number 440412/2022-6 and 408312/2023-8).

\section*{Availability and requirements}

 \textbf{Project name}: GRAMEP - Genome vaRiation Analysis from the Maximum Entropy Principle \\
 \textbf{Project home page:} \url{https://github.com/omatheuspimenta/GRAMEP} \\
 \textbf{Operating system(s):} Linux \\
 \textbf{Programming language:} Python and Rust \\
 \textbf{Other requirements:} Python $3.12$ or higher \\
 \textbf{License:} MIT \\ 
 \textbf{Any restrictions to use by non-academics:} None
 
\section*{List of abbreviations}

 SNVs: single nucleotide variants \\
 SNPs: single nucleotide polymorphisms \\
 HIV-1: human immunodeficiency virus type 1 \\
 DENV: Dengue virus  \\
 GRAMEP: Genome vaRiation Analysis from the Maximum Entropy \\

\begin{appendices}

\section{Supplementary Information}\label{secA1}

\subsection{Parameters used}\label{ap:parameters}

The parameter selection for the scenarios analyzed was conducted through empirical simulations across two distinct scenarios aimed at fine-tuning the parameters. In the first scenario, we used \textit{in silico} data from HIV, where the k-mer size was varied from 9 to 30 and applied across 100 sets of 1,000 sequences each. Performance metrics were then extracted for each simulation, including the true positive rate (TPR), false positive rate (FPR), true negative rate (TNR), false negative rate (FNR), accuracy (ACC), Matthews correlation coefficient (MCC), and F1-score. 

\begin{figure}[!h]
\centering%
\includegraphics[width=1\linewidth]{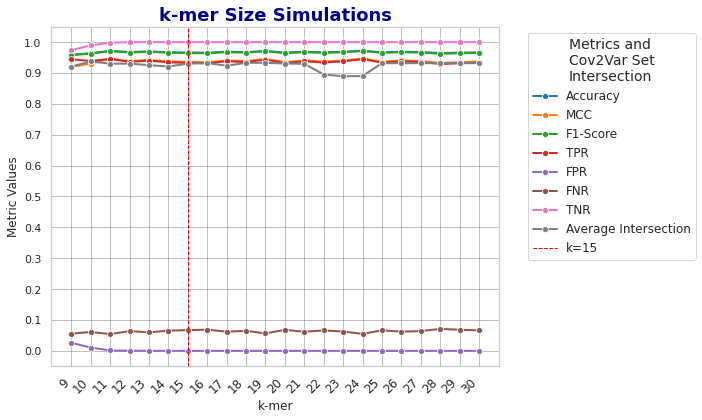}
\caption{Plot of Performance Metrics and CoV2Var set intersection for varying $k$ values}%
\label{fig:parameters_tunning}
\end{figure}

In the second scenario, which involved genomes from 20 SARS-CoV-2 variants, $10\%$ of the sequences from each variant were randomly selected. Simulations varied the k-mer size from 9 to 30, comparing the results against data provided by CoV2Var with a cutoff threshold of $95\%$, as illustrated in Figure \ref{fig:parameters_tunning}.

The analysis of these results, coupled with the principle of Occam's Razor, enabled the determination of the optimal k-mer size, which was identified as $15$. This value was selected to balance simplicity and accuracy in the method’s performance.

\subsection{GRAMEP outputs}

The `get-mutations' function generates multiple output files and a detailed report if the user opts for its creation. These output files cover various aspects of the method, providing flexibility for different applications. In total, up to 11 files can be generated if graph generation is possible. Below is a description of each file:

\begin{enumerate}
    \item \textbf{\textless variant\textgreater\_ExclusiveKmers.txt:} a plain text file containing the k-mers unique to the variant relative to the reference sequence.
    \item \textbf{\textless variant\textgreater\_ExclusiveKmers.sav:} a binary pickle version of the previous file, suitable for use in Python analyses and program routines.
    \item \textbf{\textless variant\textgreater\_FreqExclusiveKmers.csv:} a tabular file listing the identified mutations and their occurrence frequencies. A sample header and rows of the file are shown in Table \ref{tab:output1}.

    \begin{table}[!ht]
    \centering
    \caption{\textbf{\textless variant\textgreater\_FreqExclusiveKmers.csv} example}
    \label{tab:output1}
    \begin{tabular}{|l|l|l|l|}
    \hline
    position & reference\_value & variant\_value & frequency \\ \hline
    22995    & C               & A             & 14234     \\ \hline
    27874    & C               & T             & 14234     \\ \hline
    8986     & C               & T             & 14234     \\ \hline
    23403    & A               & G             & 14233     \\ \hline
    15451    & G               & A             & 14232     \\ \hline
    \end{tabular}
    \end{table}

    \item \textbf{\textless variant\textgreater\_IntersectionKmers.txt:} a text file containing the most informative k-mers common between the variant and the reference.
    \item \textbf{\textless variant\textgreater\_IntersectionKmers.sav:} a binary pickle version of the previous file for Python analyses.
    \item \textbf{\textless variant\textgreater\_reference.fasta:} a .fasta file containing the reference sequence with all mutations identified in the analysis.
    \item \textbf{\textless variant\textgreater\_report.csv:} a comprehensive report of all analyzed sequences, containing information such as sequence ID, mutation position, start and end of the genomic region, gene region type, mutation location, reference k-mer, variant k-mer, and observed mutation. If the user provides a .gff annotation file, this information is enriched. Table \ref{tab:output2} provides an example header and rows.

    \begin{landscape}
    \begin{table}[!h]
    \centering
    \caption{\textbf{\textless variant\textgreater\_report.csv} example}
    \label{tab:output2}
    \begin{tabular}{|l|l|l|l|l|l|l|l|l|l|}
    \hline
    \begin{tabular}[c]{@{}l@{}}sequence\\ id\end{tabular} & \begin{tabular}[c]{@{}l@{}}annotation\\ name\end{tabular} & start & end   & type & \begin{tabular}[c]{@{}l@{}}modification\\ localization in\\ reference\end{tabular} & \begin{tabular}[c]{@{}l@{}}reference\\ kmer\end{tabular} & \begin{tabular}[c]{@{}l@{}}exclusive\\ variant\\ kmer\end{tabular} & \begin{tabular}[c]{@{}l@{}}reference\\ snp\end{tabular} & \begin{tabular}[c]{@{}l@{}}variant\\ snp\end{tabular} \\ \hline
    AY-3\_5105                                             & ORF7a                                                     & 27394 & 27759 & gene & 27752                                                                              & AAAAGAAAGACAGAA                                          & AAAAGAAAGATAGAA                                                    & C                                                       & T                                                     \\ \hline
    AY-3\_5105                                             & S                                                         & 21563 & 25384 & gene & 22917                                                                              & ACCTGTATAGATTGT                                          & ACCGGTATAGATTGT                                                    & T                                                       & G                                                     \\ \hline
    AY-3\_5105                                             & ORF1ab                                                    & 266   & 21555 & gene & 3037                                                                               & GTATTGTTCTTTCTA                                          & GTATTGTTCTTTTTA                                                    & C                                                       & T                                                     \\ \hline
    AY-3\_5105                                             & ORF1ab                                                    & 266   & 21555 & gene & 14408                                                                              & CCCACCTACAAGTTT                                          & CCCACTTACAAGTTT                                                    & C                                                       & T                                                     \\ \hline
    AY-3\_5105                                             & N                                                         & 28274 & 29533 & gene & 28881                                                                              & AGGGGAACTTCTCCT                                          & ATGGGAACTTCTCCT                                                    & G                                                       & T                                                     \\ \hline
    AY-3\_5105                                             & N                                                         & 28274 & 29533 & gene & 28916                                                                              & CAATGGCGGTGATGC                                          & CAATGGCTGTGATGC                                                    & G                                                       & T                                                     \\ \hline
    AY-3\_5105                                             & ORF1ab                                                    & 266   & 21555 & gene & 14408                                                                              & TTCCCACCTACAAGT                                          & TTCCCACTTACAAGT                                                    & C                                                       & T                                                     \\ \hline
    \end{tabular}
    \end{table}
    \end{landscape}

    \item \textbf{\textless variant\textgreater\_variations.bed3:} a .bed3 file containing the locations of all mutations identified during the analysis.
    \item \textbf{\textless variant\textgreater\_variations.txt:} a text file listing all mutations in `location:refvar` format.
    \item \textbf{\textless variant\textgreater\_variations.sav:} a binary pickle version of `\textless variant\textgreater\_variations.txt', for use in Python.
    \item \textbf{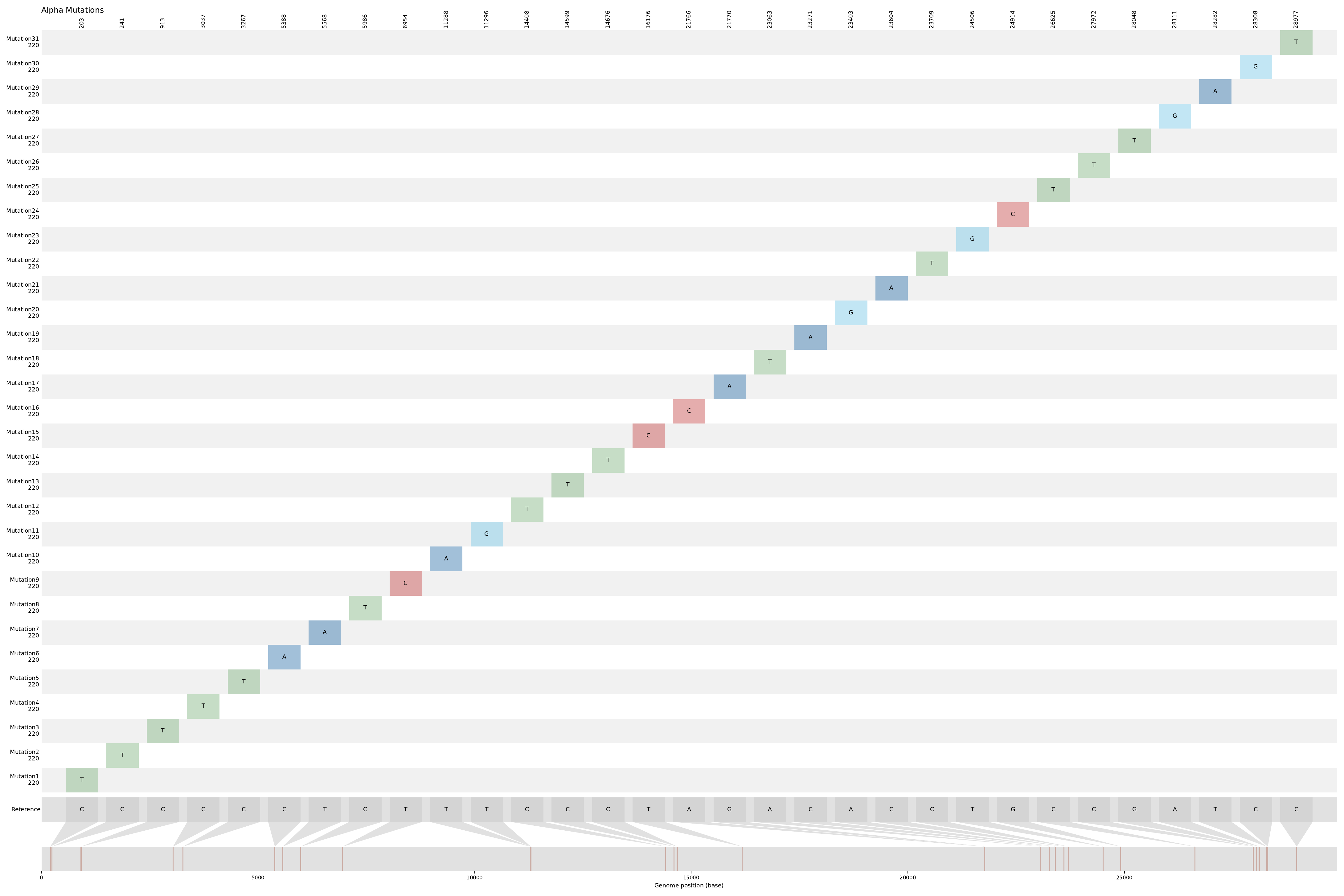:} a generated graph, based in \cite{aineniamh}, where possible, displaying the reference genome on the x-axis and the detected mutations along with their occurrence frequencies on the y-axis. Figure \ref{fig:output3} illustrates a sample graph.

    \begin{figure}[h]
    \centering%
    \includegraphics[width=1\linewidth]{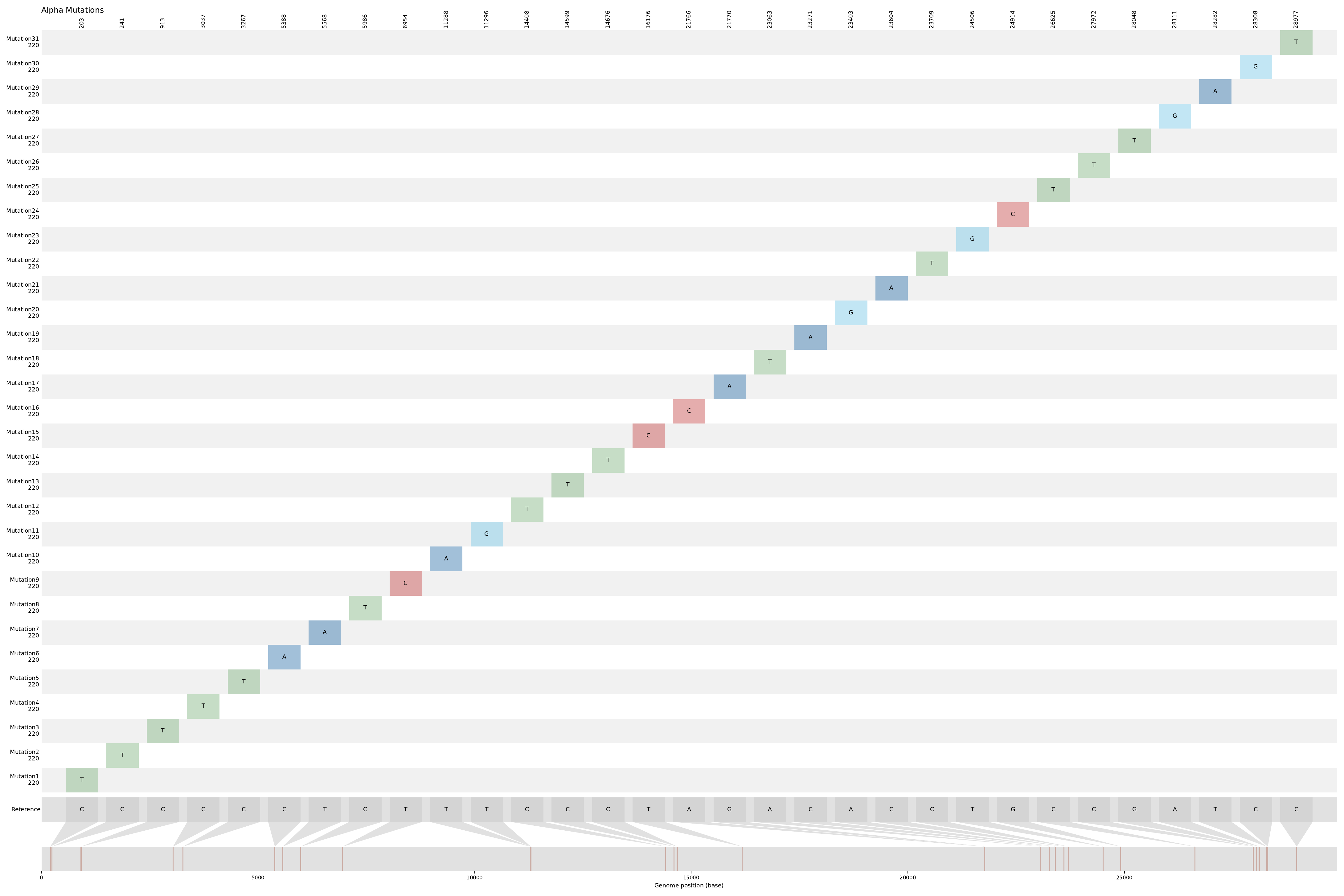}
    \caption{\textbf{results.pdf} example.}%
    \label{fig:output3}
    \end{figure}
\end{enumerate}

    The `get-intersection' function outputs two files: a text file listing all mutations shared among the selected variants and an upset plot graph illustrating the resulting sets.

    The `classify' function generates two binary files: one containing the trained model and the other containing the intervals for applying the MinMax rescaling. Additionally, three other files are provided: one contains the confusion matrix obtained during the model's training, another includes the metrics derived from the training process, and the final file holds the feature matrix extracted during model training.
\end{appendices}

\bibliography{references}% common bib file

\end{document}